# Breaking Sensitivity Barriers in Luminescence Thermometry: Synergy Between Structural Phase Transition and Luminescence Thermal Quenching

M. Tahir Abbas[1], M. Szymczak[1], M. Drozd[1], D. Szymanski[1], A. Owczarek[2], A.Musialek[1], L. Marciniak[1*]

[1] Institute of Low Temperature and Structure Research, Polish Academy of Sciences, Okólna 2, 50-422 Wrocław, Poland

[2] University of Wrocław, Faculty of Chemistry,14 F. Joliot-Curie Street, Wroclaw 50-383, Poland

*corresponding author: l.marciniak@intibs.pl

*KEYWORDS luminescence thermometry, optical sensors, ratiometric approach*

**Abstract**

One of the key parameters determining the performance of a luminescent thermometer is its relative sensitivity. In ratiometric luminescence thermometry, high relative sensitivity to temperature variations is typically achieved when the two monitored emission bands exhibit opposite thermal monotonicity. However, realizing a thermal enhancement in the luminescence intensity of one of the emission bands remains a significant challenge. In this study, we present a novel approach that leverages the synergistic effect of two phenomena: (1) the high thermal sensitivity of $Mn^{4+}$ ion luminescence, and (2) a thermally induced structural phase transition in $LaGaO_3$, which facilitates the enhancement of the luminescence signal from $Tb^{3+}$ ions in the high-temperature phase of the host material. This dual effect not only led to an increased maximum relative sensitivity but also extended the temperature range over which the sensitivity exceeded 1% K$^{-1}$. The highest recorded sensitivity was 4.5% K$^{-1}$ at 400 K. Additionally, to the best of our knowledge, the luminescence of Mn$^{4+}$ ions in the high-temperature phase of LaGaO$_3$:Mn$^{4+}$ was observed and reported here for the first time. The thermally induced modifications in the emission profile of LaGaO$_3$:Mn$^{4+}$,Tb$^{3+}$ enabled the development of a quadruple ratiometric luminescence thermometer, with complementary operating ranges, offering enhanced versatility and accuracy across a broad temperature span.

**Introduction**

Luminescence thermometry has garnered significant attention in recent years due to its distinctive features, such as rapid response, remote measurement, and ability to operate in extreme and harsh environments[1–3]. These benefits significantly addressed the limitations associated with conventional contact thermometers and facilitated remote thermal mapping with high spatial resolution and reliability, demonstrating huge potential across multiple domains[4–6]. The luminescence thermometry can be carried out by monitoring the temperature-sensitive spectroscopic parameters such as luminescence intensity ratio, luminescence kinetics, absolute emission intensity, spectral position, and emission band shape[7,8]. Among these techniques, the luminescence intensity ratio is widely regarded as the most reliable because of its independence from external interference, immunity to spectral losses, and resilience to changes in excitation density[9,10]. The most commonly described in the literature type of ratiometric luminescence thermometer relies on thermally coupled energy levels of lanthanide ions[6,11–13]. Although this approach offers very reliable temperature readings and allows the development of so called primary luminescence thermometers, its significant limitation is the relative sensitivity[6,14]. Its value is proportional to the energy difference between coupled levels[6,13]. However, the energy gap of thermally coupled ranges from 200 to 2000 cm$^{-1}$, which limits the enhancement of relative sensitivity[13,15].

Therefore, other approaches has to be explored if higher sensitivity is required. Among different ideas presented in the literature the combination of the ratio of lanthanides and transition metal ions with distinct thermal quenching behavior were introduced to overcome the disadvantages of thermally coupled-based luminescence thermometers[16–19]. Given that thermal depopulation of the emitting level in transition metal (TM) ions occurs through nonradiative electron transitions to the ground state via the intersection point of the excited and ground state potential energy parabolas, the luminescence of TM-doped phosphors is highly susceptible to thermal quenching[20–22]. This intrinsic sensitivity to temperature is a consequence of the relatively small energy gaps between electronic states and the significant electron-phonon coupling characteristic for TM ions. Furthermore, because the energies of their emitting states are strongly influenced by the crystal field strength of the host lattice, the thermal stability of TM-based phosphors can be tuned by modifying the host composition to alter the local crystal field environment[23,24]. In contrast, incorporating lanthanide ($Ln^{3+}$) ions as co-dopants enables the introduction of a reference emission signal suitable for ratiometric luminescence thermometry[25–31]. Lanthanide ions are especially advantageous in this context due to the fact that their emitting states are predominantly depopulated via multiphonon relaxation, the probability of which decreases exponentially with increasing energy gap to the next lower-lying level[32–34]. Therefore, ions such as $Eu^{3+}$ and $Tb^{3+}$, whose emitting states ($^5D_0$ and $^5D_4$, respectively) are separated from the next lower-lying levels by more than 10,000 $cm^{-1}$, are particularly well-suited for use in high-sensitivity luminescent thermometry. This approach can yield high relative sensitivities as it was proven in previous works[24,35]. However, further enhancement in sensitivity can be achieved using a strategy in which the lanthanide ion luminescence will be thermally enhanced and thus varies monotonically with temperature in the opposite direction to that of the TM ion. Although thermal enhancement of lanthanide emission intensity is uncommon, recent studies have demonstrated a mechanism whereby a temperature-induced structural phase transition in the host material alters the symmetry of the crystallographic site occupied by the dopant[36–43]. This symmetry change can significantly influence the radiative properties of the lanthanide ion, particularly in terms of transition probabilities and emission intensities.

In this context, lanthanide ions have hitherto been exclusively employed as dopants to exploit such symmetry-driven changes in spectroscopic behavior[39,42,43]. Within the temperature range of the phase transition, a pronounced increase in emission intensity is observed for the high-temperature (HT) phase of the phosphor, while a concurrent decrease is noted for the low-temperature (LT) phase. The thermal enhancement of luminescence in the HT phase can be particularly advantageous for improving the relative sensitivity of ratiometric luminescent thermometers, as it creates a strong and distinct thermometric signal that can be used for precise temperature sensing.

In this study, we propose a novel ratiometric luminescence thermometry approach, schematically illustrated in Figure 1. The strategy is based on the dual doping of a host material with $Mn^{4+}$ and $Tb^{3+}$ ions representative dopants from TM and lanthanide $Ln^{3+}$ groups, respectively. $Mn^{4+}$ ions, with a well-characterized *3d³* electronic configuration, serve as the thermally sensitive probe, whereas $Tb^{3+}$ ions are selected for their favorable spectroscopic properties, particularly the large energy gap (>10,000 cm$^{-1}$) between the emitting $^5D_4$ level and the lower-lying $^7F_0$ level. Although $Eu^{3+}$ ions exhibit a similarly large energy separation, their emission bands spectrally overlap significantly with those of $Mn^{4+}$, making it difficult to spectrally resolve their respective luminescence signals, and thus complicating accurate temperature readout.

In conventional ratiometric thermometry systems based on intensity ratios between TM and $Ln^{3+}$ emissions under selective excitation, it is typically observed that the TM emission decreases with increasing temperature due to thermal quenching, whereas the $Ln^{3+}$ emission remains thermally stable or exhibits only a slight decrease. This contrasting behavior allows for a high degree of thermal modulation in the luminescence intensity ratio (*LIR*), which translates to high relative sensitivity ($S_R$). However, in host materials undergoing thermally induced structural phase transitions, an enhancement of the $Ln^{3+}$ emission intensity with increasing temperature can be observed, as previously reported[37,42,44]. This thermally induced increase in luminescence from $Ln^{3+}$ ions offers an opportunity for even greater LIR dynamics, thereby enabling the achievement of significantly higher relative sensitivities. In this work, $LaGaO_3$ was selected as the host material due to its unique combination of favorable properties: (1) it undergoes a structural phase transition from orthorhombic to rhombohedral at elevated temperatures[45–47]; (2) it contains a $La^{3+}$ crystallographic site which can be replaced by the lanthanide ions such as $Tb^{3+}$; and (3) it features $Ga^{3+}$ octahedral sites suitable for hosting $Mn^{4+}$ ions, which are known to occupy such coordination environments to exhibit luminescence.

Notably, to the best of our knowledge, this is the first study to investigate and report in details the influence of the $LaGaO_3$ structural phase transition on the luminescent properties of $Mn^{4+}$ ions. The thermally driven modifications in the luminescence spectra of $LaGaO_3$:$Mn^{4+}$,$Tb^{3+}$ enable the implementation of multiple ratiometric thermometry strategies. These complementary approaches allow for an extended operational temperature range while maintaining high relative sensitivity, thus enhancing the practical applicability and precision of luminescent temperature sensing based on this material system.

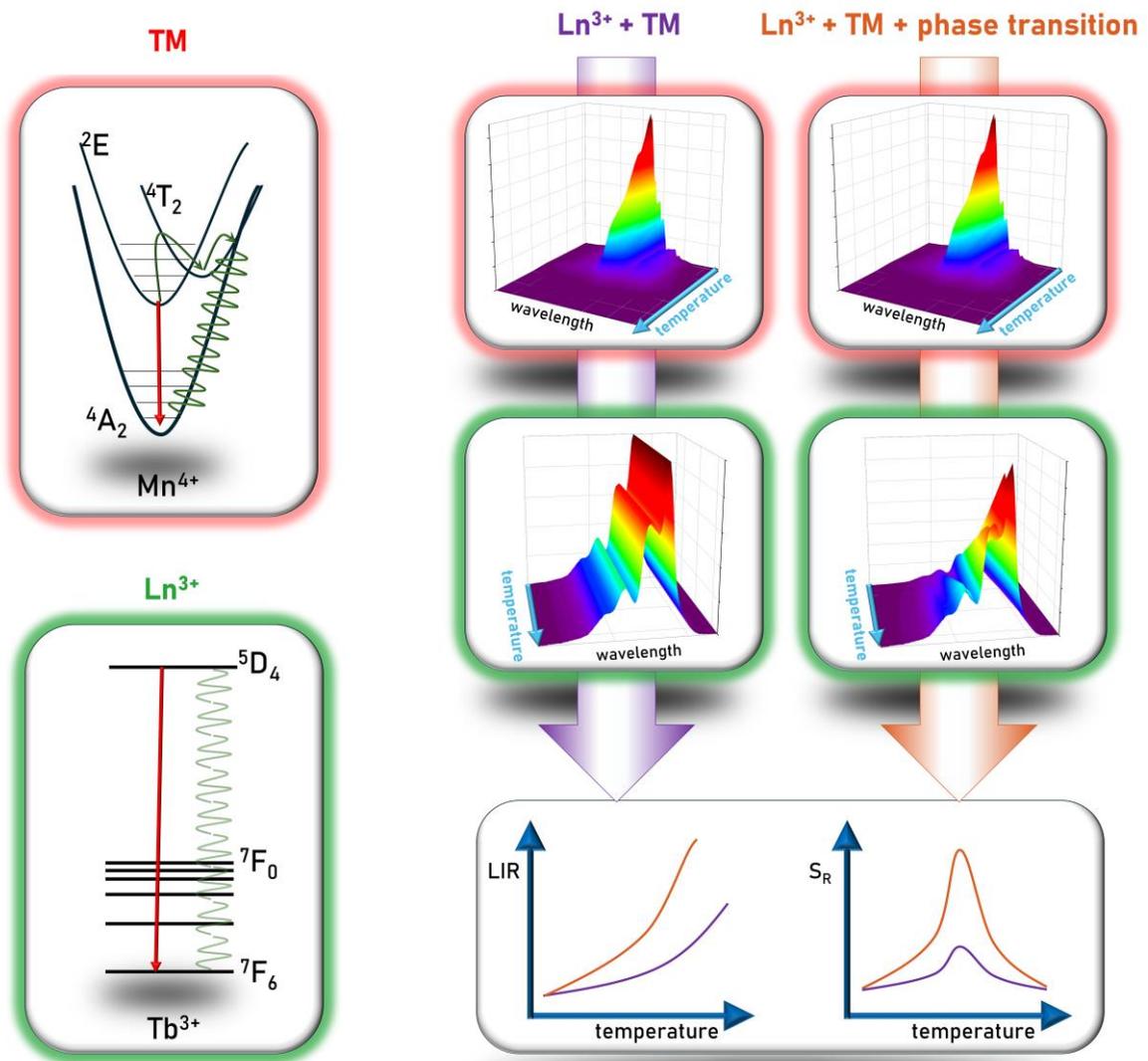

**Figure 1**. Schematic presentation of the main concept of the proposed approach: the comparison of the expected thermal changes in the emission intensity of transition metal ions (TM) and lanthanide ions ($Ln^{3+}$) in the typical ratiometric approach ($Ln^{3+}$+TM) and in the host material which undergoes of thermally induced first order phase transition ($Ln^{3+}$+ TM + phase transition).

## 2. Experimental Section

*Materials*

The $LaGaO_3$:x%$Mn^{4+}$ (where x = 0.001, 0.01, 0.05, 0.1, 0.5), $LaGaO_3$:0.25%$Tb^{3+}$, $LaGaO_3$:0.25%$Tb^{3+}$, 0.001%$Mn^{4+}$ phosphors were synthesized by conventional high-temperature solid-state reaction method. $La_2O_3$ (99.999% purity, Stanford Materials

Corporation), $Ga_2O_3$ (99.999% purity, Alfa Aesar), $MnCl_2 \cdot 4H_2O$ (99% purity, Sigma-Aldrich), $Tb_4O_7$ (99.999% purity, Standford Materials Corporation) were used as starting materials. These raw materials were stoichiometrically weighed according to the required chemical formula, then mixed and ground with a few drops of hexane in an agate mortar for 30 minutes to achieve homogeneity. The mixtures were annealed in porcelain crucibles for 3 hours in a muffle furnace at 873 K with a heating rate of 10 K min$^{-1}$. After cooling to room temperature, the obtained powders were reground and then sintered at 1673 K for 6 hours with a heating rate of 10 K min$^{-1}$. Finally, the obtained powders were ground in an agate mortar for further characterization.

*Methods*

Powder diffraction data were obtained in Bragg-Brentano geometry using a PANalytical Aeris diffractometer, which utilizes Ni-filtered Cu K$\alpha$ radiation with a wavelength of $\lambda$=1.54060 Å (V = 40 kV, I = 15 mA). Measurments were conducted at room temperature within an angular range of 2theta=5-90°, with a step of 0.0217329°.

The morphology and chemical composition of the samples were characterized by using a Field Emission Scanning Electron Microscope (FE-SEM, FEI Nova NanoSEM 230) equipped with an energy dispersive X-ray spectrometer (EDX, EDAX Apollo X Silicon Drift Correction) compatible with Genesis EDAX microanalysis Software (version 6.0). Prior to SEM imaging, the $Mn^{4+}$-doped $LaGaO_3$ sample was dispersed in alcohol, and then a drop of suspension was inserted in a carbon stub. Finally, the SEM images were captured at an accelerating voltage of 5.0 kV in a beam deceleration mode, which enhances the imaging characteristics. In contrast, EDS measurements were performed at 30 kV in order to obtain high quality EDS maps of the elements included in the $Mn^{4+}$-doped $LaGaO_3$ sample.

A differential scanning calorimetry (DSC) measurements were carried out on a Perkin-Elmer DSC 8000 calorimeter equipped with a Controlled Liquid Nitrogen Accessory LN2 with a heating and cooling rate of 20 K min$^{-1}$. The sample was sealed in an aluminum pan. The measurement for the powder sample was conducted in the temperature range from 300 K to 500 K.

The excitation and emission spectra were measured using the FLS1000 Fluorescence Spectrometer from Edinburgh Instruments equipped with a 450 W Xenon lamp and R928 photomultiplier tube from Hamamatsu. The temperature of the sample was controlled by using a THMS600 heating – cooling stage from Linkam (0.1 K temperature stability and 0.1 K point resolution) during temperature-dependent measurements. The luminescence decay profiles were also obtained using the FLS1000 equipped with a 150 W µFlash lamp. The average lifetime of the excited states was calculated with the double exponential function (Eqs.1 and 2):

$$\tau_{avr} = \frac{A_1\tau_1^2 + A_2\tau_2^2}{A_1\tau_1 + A_2\tau_2} \quad (1)$$

$$I(t) = I_0 + A_1 \cdot \exp\left(-\frac{t}{\tau_1}\right) + A_2 \cdot \exp\left(-\frac{t}{\tau_2}\right) \quad (2)$$

where $\tau_1$ and $\tau_2$ represent decay components and $A_1$ and $A_2$ are the amplitudes of double-exponential functions.

## 3. Results and discussion

At room temperature, LaGaO$_3$ crystallizes in the orthorhombic perovskite-type structure with the *Pbnm* (No. 62) space group, which is widely described in the literature[48–50]. Under specific external stimuli - such as elevated temperature or applied pressure - LaGaO$_3$ undergoes a reversible first-order phase transition to a higher-symmetry rhombohedral phase, belonging to the *R3c* (No. 167) space group. This structural transformation is accompanied by a change in the local point symmetry at the cationic sites -from $C_S$ in the orthorhombic phase to $D_3$ in the rhombohedral one - which has a profound influence on the crystal field environment and, consequently, on the luminescence behavior of dopant ions. The enhanced symmetry of the rhombohedral LaGaO$_3$ structure stems from the uniformity of Ga-O distances, all measuring 1.977 Å. In contrast, the orthorhombic structure exhibits a distortion manifested by three distinct lengths of Ga-O bonds: 1.982 Å, 1.977 Å, and 1.972 Å - arranged in pairs along specific crystallographic axes. This variation breaks the symmetry of the octahedral environment, making the orthorhombic phase less symmetric. Therefore, the rhombohedral structure is inherently more symmetric due to the regularity and equivalence of its bonding geometry.

From a crystallographic perspective, both structures of LaGaO$_3$ consist of a network of corner-sharing GaO$_6$ octahedra and La$^{3+}$ ions in distorted coordination polyhedra involving 8 to 12 oxygen atoms. The degree of octahedral tilting and distortion is significantly more pronounced in the orthorhombic phase, resulting in the reduced symmetry relative to the rhombohedral structure. LaGaO$_3$ serves as an excellent host for co-doping with transition metal (TM) and lanthanide (Ln$^{3+}$) ions, as confirmed by numerous studies. Typically, TM ions occupy the octahedrally coordinated Ga$^{3+}$ sites, while Ln$^{3+}$ ions substitute for La$^{3+}$, reflecting the favorable match in ionic radii and coordination environments. Same is in the case for the Tb$^{3+}$ and Mn$^{4+}$ co-doped LaGaO$_3$ investigated in the present study, where the dopant ions selectively substitute the La$^{3+}$ and Ga$^{3+}$ sites, respectively. The analysis of the room temperature XRD patterns for LaGaO$_3$:Mn$^{4+}$ (Figure 2b) and LaGaO$_3$:Tb$^{3+}$ and LaGaO$_3$:Mn$^{4+}$,Tb$^{3+}$ (Figure S1) confirms the structural purity of the obtained samples and the fact that the change in the dopant concentration does not significantly affect the structure of the obtained phosphors. The DSC measurements reveled the first order phase transition in LaGaO$_3$:Mn$^{4+}$ at temperature around 420 K (Figure 2c). Usually an increase in the dopant concentration due to the difference between ionic radii of dopant and substituted host material cation leads to the change in the phase transition temperature. However in the case of the LaGaO$_3$:Mn$^{4+}$ low dopant concentration used leads to only slight increase of the phase transition temperature from 420.1 K for LaGaO$_3$:0.001%Mn$^{4+}$ to 420.55 K for LaGaO$_3$:0.5%Mn$^{4+}$. In order to eliminate the energy transfer between dopant ions which may affect the thermometric performance of the luminescence thermometer the higher concentration of $Mn^{4+}$ ions were not investigated. The analysis of the Raman spectra measured for $LaGaO_3$ at 83 K and 453 K indicates a significant difference induced by the structural phase transition (Figure 2d). The low temperature Raman spectra consist of larger number of modes[50–52]. The vibrations at 100 cm$^{-1}$, 115 cm$^{-1}$, 135 cm$^{-1}$, 144 cm$^{-1}$ and 171 cm$^{-1}$ are associated with the internal $La^{3+}$ ion vibration. The vibrations at 434 cm$^{-1}$ and 450 cm$^{-1}$ are associated with the internal bending and stretching modes of the $GaO_6$ group, while the stretching motion of the B-O bond results in the occurrence of the vibrations in the 600-780 cm$^{-1}$ range[51]. On the other hand the increase in the structure symmetry induced by the phase transition leads to the reduction of the peaks in the Raman spectra of $LaGaO_3$. The rearrangement of the $La^{3+}$ cations associated with the displacement of the $Ga^{3+}$ cations with the change of the tilt of the $GaO_6$ group results in the observation only several bands. Bands below 200 cm$^{-1}$ and at 259 cm$^{-1}$ can be assign to the internal La vibrations and pure oxygen bending vibration, respectively. The reduction in the number of lines associated with the vibrations of the B-O bond and $GaO_6$ group can be also found. The fact that the observed change in the Raman spectra is strictly associated with the phase transition of the $LaGaO_3$ was confirmed based on the Raman spectra measured as a function of temperature (Figure 2e). The sharp change in the shape of the spectra at around 420 K correlated with the phase transition temperature obtained based on the DSC analysis. The morphological studies performed using SEM technique revealed that the obtained phosphors consists of high aggregated crystals with around 4 μm in average diameter (Figure 2f). The homogeneous elemental distribution in the synthesized $LaGaO_3$:$Mn^{4+}$ was also confirmed by the EDS analysis (Figure 2g -j).

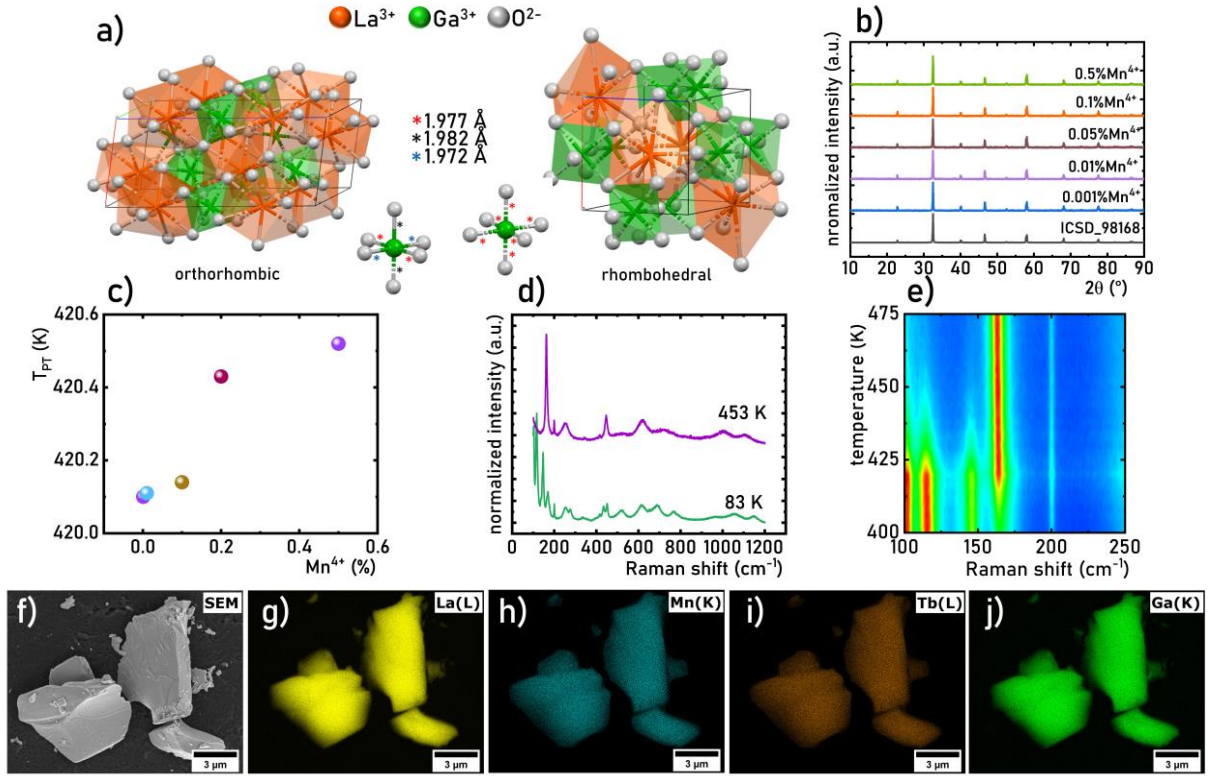

**Figure 2**. Visualization of the orthorhombic and rhombohedral structures of LaGaO$_3$ -a); The comparison of the room temperature XRD patters of LaGaO$_3$:Mn$^{4+}$ with different concentration of Mn$^{4+}$ ions -b); the influence of the Mn$^{4+}$ concentration on the phase transition temperature ($T_{PT}$) for LaGaO$_3$:Mn$^{4+}$ obtained from the DSC measurements-c); the Raman spectra of LaGaO$_3$ measured at 83K and 453 K-d); thermal dependence of the normalized Raman spectra of LaGaO$_3$-e); representative SEM image of LaGaO$_3$:Mn$^{4+}$ -f) and corresponding elemental distribution for La (yellow)-g), Mn (cyan)-h), Tb (orange)-i) and Ga (green)-j).

As previously discussed, the luminescence spectrum of Mn$^{4+}$-doped phosphors is characterized by an emission band centered around 690 nm, corresponding to the spin-forbidden $^2E \rightarrow {}^4A_2$ electronic transition[20,22,53,54]. Due to the weak electron-phonon coupling of the $^2E$ excited state, there is minimal displacement between the energy levels parabolas of the ground $^4A_2$ state and the $^2E$ state in the wavevector domain. Consequently, the $^2E \rightarrow {}^4A_2$ emission band is spectrally narrow and typically features both a zero-phonon line (ZPL) and accompanying phonon sidebands[53]. In the case of LaGaO$_3$:Mn$^{4+}$ at 83 K, the ZPL is observed at 701 nm, while the emission is predominantly governed by a phonon sideband centered at 711 nm (Figure 3a). It was found that the $Mn^{4+}$ concentration up to 0.5%$Mn^{4+}$ does not affect the shape of the emission spectra of $LaGaO_3:Mn^{4+}$ measured at 83 K (Figure 3a). Although the Tanabe-Sugano diagram for $3d^3$ electronic configurations indicates that the energy of the $^2E$ state is largely independent of crystal field strength, changes in this parameter significantly influence the spectroscopic behavior of $Mn^{4+}$ ions[54–56]. Specifically, an increase in crystal field strength raises the energy of the $^4T_2$ level, which intersects both the $^2E$ and $^4A_2$ levels. These intersections have implications for the thermal stability of luminescence from the $^2E$ state. Furthermore, due to the partial wavefunction overlap between the $^2E$ and $^4T_2$ states, and the difference in their transition selection rules ($^2E \rightarrow {}^4A_2$ being spin-forbidden and $^4T_2 \rightarrow {}^4A_2$ being spin-allowed), the energy gap between these levels plays a crucial role in governing the luminescence kinetics of the $^2E$ state[57–59]. As a result, weaker crystal fields are generally associated with shorter $^2E$ state lifetimes[21]. In the case of the $LaGaO_3:Mn^{4+}$, a slight reduction in emission intensity is observed with increasing temperature up to approximately 300 K. An increase in the temperature results in an increase in the anti-Stokes part of the emission band of $Mn^{4+}$ which is caused by the thermalization of the vibrational states (Figure 3b, Figure S2-3). However, for $LaGaO_3:0.001\%Mn^{4+}$ above around 420 K an additional band occurs at 682 nm and becomes dominant in the $LaGaO_3:Mn^{4+}$ spectra at higher temperatures (Figure 3c and 3d). Since the temperature above which this band is observed correlates well with the phase transition temperature in the analyzed material, this emission band was assigned to the anti-Stokes part of the $^2E \rightarrow {}^4A_2$ emission band of the high temperature phase of $LaGaO_3:Mn^{4+}$. It is worth to notice that this band was not observed for higher $Mn^{4+}$ concentration, which may suggest that the interionic energy transfer may lead to its quenching (Figure S4, S5). Excitation spectra recorded below the phase transition temperature reveal two distinct bands, with maxima at approximately 366 nm and 511 nm, corresponding to the $^4A_2 \rightarrow {}^4T_1$ and $^4A_2 \rightarrow {}^4T_2$ transitions, respectively (Figure S6). Above the phase transition temperature, these bands exhibit a slight red shift, with new maxima appearing at 376 nm and 520 nm, respectively. This spectral shift is attributed to a change in the crystal field strength associated with the structural phase transition, varying from $Dq/B$=2.47 to 2.64. Additionally the covalency of the $Mn^{4+}$-$O^{2-}$ bond changes due to this phase transition from $\beta_1$= 0.948 to 0.931. Analysis of the temperature-dependent integrated emission intensity of $Mn^{4+}$ ions reveals minimal variation up to ~300 K, regardless of dopant concentration (Figure 3e). Beyond this point, a sharp decline is observed, corresponding to the thermally activated nonradiative depopulation of the $^2E$ state. Notably, the onset of thermal quenching occurs below the phase transition temperature, indicating that this effect arises from thermally activated crossing between the $^2E$ and $^4T_2$ parabolas rather than from the structural transformation of the host lattice. From this temperature dependence, the activation energy for nonradiative depopulation of the $^2E$ state in $LaGaO_3$:$Mn^{4+}$ was determined to be $E_a = 0.52$ eV (Figure S7). Thermally induced spectral changes, particularly in the regions 682 - 684 nm (HT phase) and 711 - 713 nm (LT phase), exhibit opposite trends with temperature (Figure 3f), enabling the development of a ratiometric luminescence thermometer using the luminescence intensity ratio (*LIR*) as a thermometric parameter:

$$LIR_1 = \frac{\int_{681nm}^{683nm} \left(^2E \rightarrow {}^4A_2\right) d\lambda}{\int_{711nm}^{714nm} \left(^2E \rightarrow {}^4A_2\right) d\lambda} \quad (3)$$

The temperature dependence of $LIR_1$ demonstrates a gradual increase up to $T_{PT}$, followed by a rapid, nearly 3-fold increase (for 0.001% $Mn^{4+}$) beyond this temperature. Further temperature elevation results in the reduction in the $LIR_1$ value, due to the quenching of the intensity of both analyzed signals (Figure 3g). To quantify the thermometric performance of this ratiometric luminescence thermometer, the relative sensitivity ($S_R$) was calculated as follows:

$$S_R = \frac{1}{LIR}\frac{\Delta LIR}{\Delta T}\cdot 100\% \qquad (4)$$

where $\Delta LIR$ represents the change of $LIR$ corresponding to the change in temperature by $\Delta T$. The temperature dependence of $S_R$ decreases with temperature up to around 350 K above which a rapid increase in the intensity is observed with the maxima reaching $S_R$= 2.7% K$^{-1}$ at 415 K.

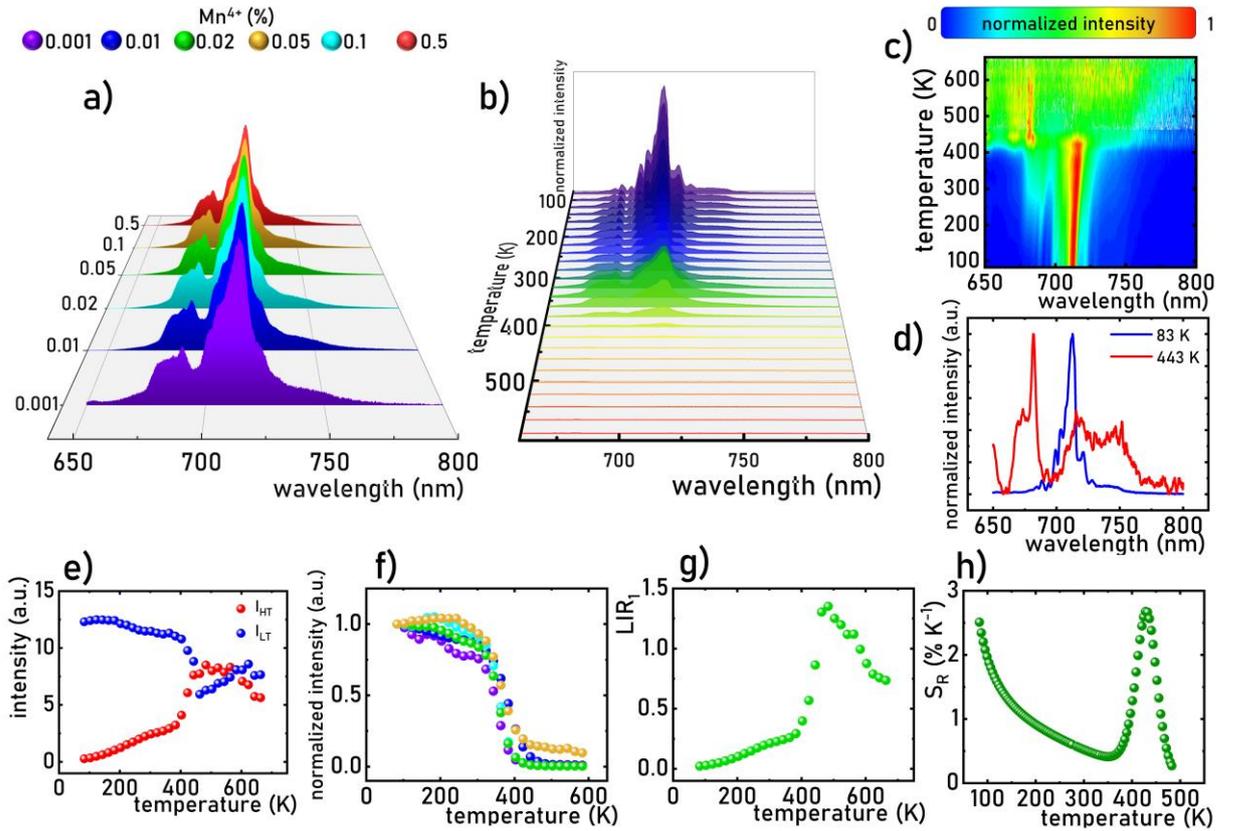

**Figure 3.** Comparison of the emission spectra of LaGaO$_3$:Mn$^{4+}$ measured at 83 K for different Mn$^{4+}$ concentrations ($\lambda_{exc}$ = 377 nm)-a); emission spectra of LaGaO$_3$:0.001%Mn$^{4+}$ measured as a function of temperature-b) thermal luminescence maps of normalized emission spectra of LaGaO$_3$:0.001%Mn$^{4+}$ -c); the comparison of the normalized emission spectra of LaGaO$_3$:0.001%Mn$^{4+}$ measured at 83 K and 443 K-d); the influence of temperature on the integral emission intensity of Mn$^{4+}$ ions in LaGaO$_3$:Mn$^{4+}$-e); thermal dependence of the emission intensity of the LT (integrated in the 711-713 nm spectral range) and HT (integrated in the 681 - 683 nm spectral range) phases of LaGaO$_3$:0.001%Mn$^{4+}$ -f); thermal dependence of $LIR_1$ -g) and corresponding $S_R$ -h).

The luminescent properties of $Tb^{3+}$ ions are primarily attributed to radiative transitions from the excited $^5D_4$ level to the lower $^7F_J$ multiplet levels. These transitions result in characteristic emission bands at approximately 490 nm, 545 nm, 580 nm, and 620 nm, corresponding to the $^5D_4 \rightarrow {}^7F_6$, $^5D_4 \rightarrow {}^7F_5$, $^5D_4 \rightarrow {}^7F_4$, and $^5D_4 \rightarrow {}^7F_3$ electronic transitions, respectively. As illustrated in Figure 4a, in the case of $LaGaO_3:Tb^{3+}$, additional, less intense emission bands are observed in the 400-470 nm range (Figure 4a). These are associated with $^5D_3 \rightarrow {}^7F_J$ transitions, which are rarely detected in $Tb^{3+}$-doped phosphors due to the relatively small energy gap (~5000 cm$^{-1}$) between the $^5D_3$ and $^5D_4$ levels. This narrow gap facilitates efficient multiphonon nonradiative relaxation from the $^5D_3$ level, especially at elevated temperatures, resulting in rapid thermal quenching of the corresponding emission. In contrast, emissions originating from the $^5D_4$ level exhibit significantly greater thermal stability. Nevertheless, the luminescence intensity of $LaGaO_3:Tb^{3+}$ still decreases with increasing temperature. A detailed temperature-dependent analysis of the $LaGaO_3:Tb^{3+}$ emission spectra reveals a noticeable spectral shift in the emission maxima near the phase transition temperature of the $LaGaO_3$ host. This effect is particularly pronounced for the $^5D_4 \rightarrow {}^7F_6$ band, where the emission peak observed at ~491 nm for the LT phase of $LaGaO_3:Tb^{3+}$ shifts to ~485 nm above 420 K, corresponding to the HT phase. In the case of the $^5D_4 \rightarrow {}^7F_5$ transition, a reduction in the number of observable Stark components from 5 to 4 is detected above the phase transition temperature, along with a slight shift in the Stark line maximum. A similar spectral shift is also evident in the $^5D_4 \rightarrow {}^7F_4$ transition band under these conditions. These effects are associated with the symmetry increase of the crystallographic site occupied by the $Tb^{3+}$ ions. Analysis of the thermal dependence of emission intensities for the LT and HT phases of $LaGaO_3:Tb^{3+}$ demonstrates that the integral luminescence intensity of the LT phase remains relatively constant up to the phase transition temperature, above which a rapid intensity decrease occurs. Conversely, the emission lines associated with the HT phase exhibit an almost threefold increase in intensity above the transition point, with saturation occurring around 500 K. This opposing thermal behavior of the emission intensity of $Tb^{3+}$ ions in LT and HT phase can be exploited for ratiometric luminescence thermometry using LIR's defined as follows:

$$LIR_2 = \frac{\int_{586nm}^{588nm} \left(^5D_4 \to {}^7F_4\right) d\lambda}{\int_{491nm}^{493nm} \left(^5D_4 \to {}^7F_6\right) d\lambda} \quad (5)$$

$$LIR_3 = \frac{\int_{487nm}^{489nm} \left(^5D_4 \to {}^7F_6\right) d\lambda}{\int_{491nm}^{493nm} \left(^5D_4 \to {}^7F_6\right) d\lambda} \quad (6)$$

Both $LIR_2$ and $LIR_3$ show a moderate increase with temperature, followed by a pronounced threefold rise upon exceeding the phase transition temperature. Above 500 K, $LIR_2$ and $LIR_3$ become temperature-independent, reflecting the thermal stability of emissions from both LT and HT phases in this regime. The derived $S_R$ aligns with the trends observed in $LIR_2$ and $LIR_3$, with $S_R$ values exceeding 0.5% $K^{-1}$ within the temperature range corresponding to the structural phase transition of $LaGaO_3:Tb^{3+}$. In both cases, maximum sensitivities of $S_R$=2.8% $K^{-1}$ are achieved near 460 K. The concentration of $Tb^{3+}$ ions employed in this study was selected to ensure high luminescence intensity. However, based on prior studies, an increase in $Tb^{3+}$ ion concentration due to the ionic radius mismatch between $Tb^{3+}$ and $La^{3+}$-would be expected to result in an elevation of the phase transition temperature, as previously observed for $LaGaO_3:Eu^{3+}$ [60].

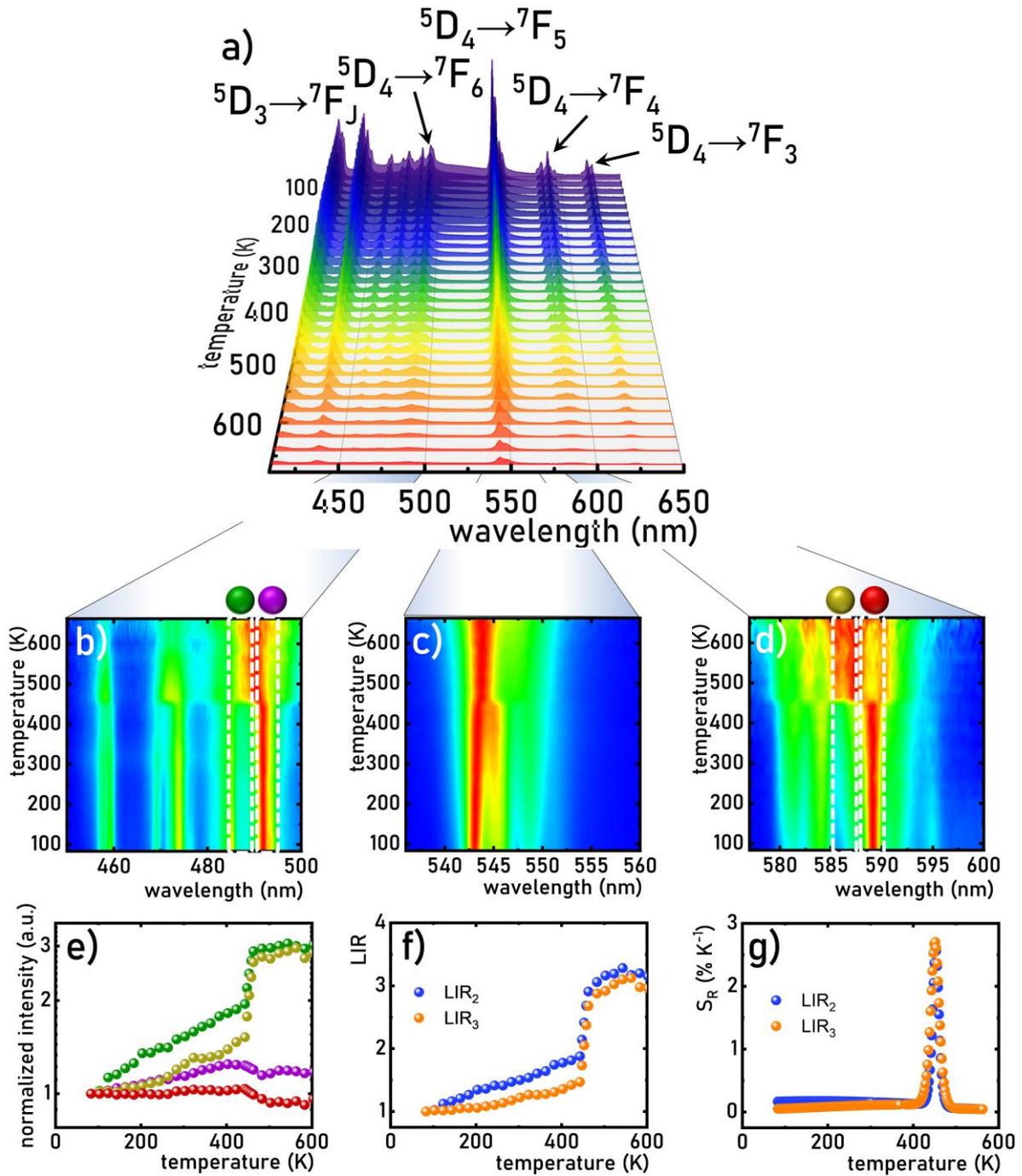

**Figure 4.** Emission spectra of LaGaO$_3$:Tb$^{3+}$ measured as a function of temperature ($\lambda_{exc}$ = 377 nm) -a); thermal luminescence maps of normalized emission spectra of LaGaO$_3$:Tb$^{3+}$ presented in the spectra range corresponding to the $^5D_3 \rightarrow {}^7F_J$ + $^5D_4 \rightarrow {}^7F_6$ - b); $^5D_4 \rightarrow {}^7F_5$ – c) and $^5D_4 \rightarrow {}^7F_4$ – d); the influence of the temperature on the emission intensities integrated in the spectral ranges presented in Figures 4b and d -e); thermal dependencies of $LIR_2$ and $LIR_3$ -f) and corresponding $S_R$-g).

The high temperature sensitivity of $Mn^{4+}$ and $Tb^{3+}$ ion luminescence in $LaGaO_3$ makes this host material highly suitable for ratiometric luminescence thermometry. To explore this potential, $LaGaO_3$ co-doped with 0.001% $Mn^{4+}$ and 0.25% $Tb^{3+}$ was synthesized, and its temperature-dependent luminescence spectra were measured and are presented in Figure 5a. At 83 K, the emission spectrum of $LaGaO_3$:$Mn^{4+}$,$Tb^{3+}$ is dominated by the $^2E \rightarrow ^4A_2$ transition of $Mn^{4+}$ ions, although the characteristic emission bands of $Tb^{3+}$ ions are also distinctly visible. Excitation spectra measured while monitoring the emission of $Tb^{3+}$ ($\lambda_{em}$ = 545 nm) and $Mn^{4+}$ ($\lambda_{em}$ = 711 nm) confirm the absence of energy transfer between the two dopant species, which is primarily attributed to their low concentrations (Figure S8). Upon increasing temperature, a gradual decline in the luminescence intensity of $Mn^{4+}$ is observed, followed by a sharp quenching above 300 K. Nevertheless, a residual $Mn^{4+}$ emission remains detectable even beyond 500 K. In contrast, the $Tb^{3+}$ emission bands persist throughout the entire investigated temperature range. The evolving ratio of $Mn^{4+}$ (red emission) to $Tb^{3+}$ (green emission) intensities with temperature results in a progressive shift in the color of the emitted light. This is evidenced by the transformation of chromaticity coordinates (CIE 1931): from an orange-red at 83 K, through yellow and white, to a greenish-blue at 583 K, where the spectrum is predominantly governed by $Tb^{3+}$ emission. Chromaticity analysis indicates that these changes are primarily driven by a variation in the *x*-coordinate, with the most significant shifts occurring between 300 K and 500 K, coinciding with the thermal quenching region of $Mn^{4+}$ emission. The integral luminescence intensity of $Mn^{4+}$ in $LaGaO_3$:$Mn^{4+}$,$Tb^{3+}$ presents the similar behavior previously described for $LaGaO_3$:$Mn^{4+}$, further supporting the conclusion that there is no significant interaction or energy transfer between the $Mn^{4+}$ and $Tb^{3+}$ ions in the host material. Notably, the total emission intensity of $Tb^{3+}$ also decreases slightly with increasing temperature, although this reduction is far less pronounced than that observed for $Mn^{4+}$. Interestingly, the $Tb^{3+}$ emission component associated with the HT phase of $LaGaO_3$ increases steadily up to approximately 400 K, after which a sharp intensity enhancement occurs. Above 500 K, the HT-phase-related $Tb^{3+}$ emission intensity stabilizes, showing minimal further temperature dependence. The contrasting thermal responses of $Mn^{4+}$ and $Tb^{3+}$ luminescence (both total and HT-phase-specific) enable the definition of two distinct ratiometric parameters, $LIR_4$ and $LIR_5$.

$$LIR_4 = \frac{\int_{530nm}^{590nm} \left(^5D_4 \to {}^7F_4\right)[Tb^{3+}]d\lambda}{\int_{670nm}^{750nm} \left(^2E \to {}^4A_2\right)[Mn^{4+}]d\lambda} \quad (7)$$

$$LIR_5 = \frac{\int_{586nm}^{588nm} \left(^5D_4 \to {}^7F_4\right)[Tb^{3+}]d\lambda}{\int_{670nm}^{750nm} \left(^2E \to {}^4A_2\right)[Mn^{4+}]d\lambda} \quad (8)$$

$LIR_4$ corresponds to a conventional ratiometric approach, widely reported in the literature, and is based on the intensity ratio between thermally quenched $Mn^{4+}$ emission and relatively stable $Tb^{3+}$ emission. A significant increase in $LIR_4$ is observed in the temperature range where $Mn^{4+}$ emission undergoes efficient quenching. $LIR_5$, on the other hand, leverages the thermally enhanced emission of $Tb^{3+}$ ions in the HT phase of $LaGaO_3$. This parameter exhibits a remarkably strong thermal dependence, increasing by more than 900-fold relative to its value at 83 K, with the most significant changes occurring above 300 K. The $S_R$ profiles derived from both $LIR_4$ and $LIR_5$ reveal important distinctions. In the case of $LIR_4$, the maximum $S_R$ reaches 3.2% $K^{-1}$ at 420 K, with high sensitivity observed over a relatively narrow range (approximately 350 - 480 K). In contrast, $LIR_5$ achieves a substantially higher maximum $S_R$ of 4.7% $K^{-1}$, and the temperature range over which high sensitivity is maintained is significantly broader (approximately 220 - 500 K). This extended range results from the synergistic interplay of two key effects: the thermal quenching of $Mn^{4+}$ luminescence and the phase transition in $LaGaO_3$, which enhances the $Tb^{3+}$ emission associated with the HT phase. This combination of phenomena provides direct and compelling evidence for the efficacy of the dual-dopant, phase-transition-based thermometric approach proposed in this work.

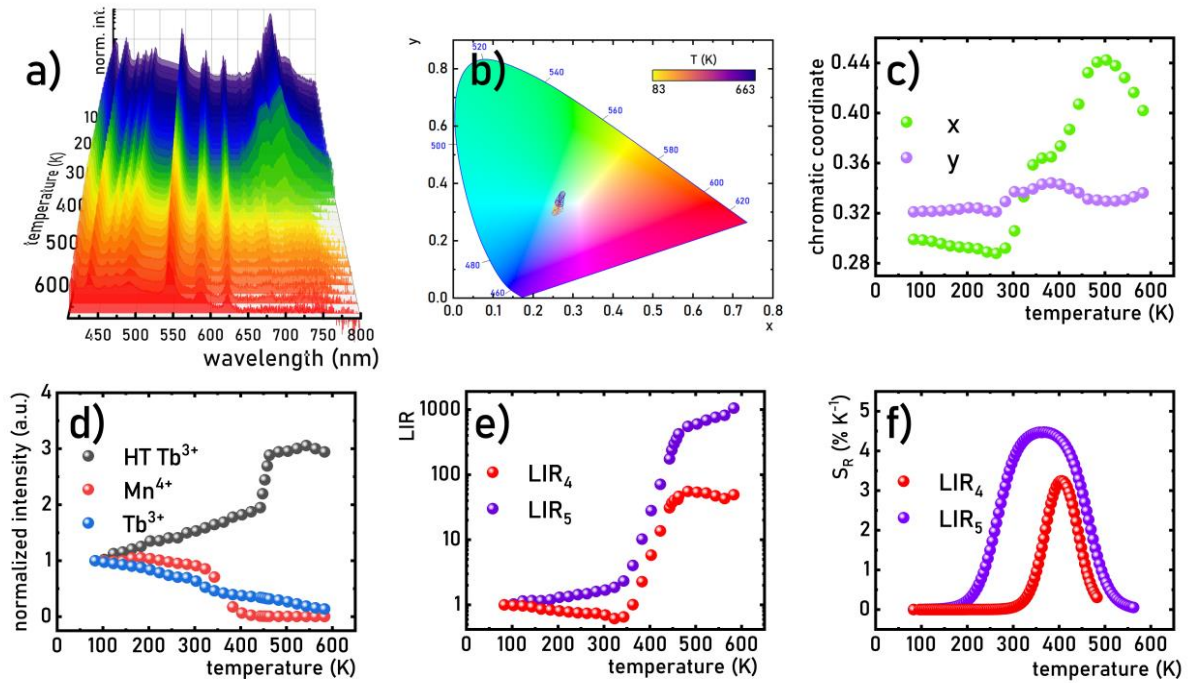

**Figure 5.** Emission spectra of $LaGaO_3$:0.001%$Mn^{4+}$, 0.25%$Tb^{3+}$ measured as a function of temperature - a); thermal dependence of the chromatic coordinates of the light emitted by the $LaGaO_3$:$Mn^{4+}$, $Tb^{3+}$ presented as a CIE 1931 diagram -b) and their values as a function of temperature – c); the influence of the temperature on the integral emission intensities of $Mn^{4+}$ ions, $Tb^{3+}$ ions and $Tb^{3+}$ ions in high temperature phase of $LaGaO_3$:0.001%$Mn^{4+}$, 0.25%$Tb^{3+}$ (HT $Tb^{3+}$) -d); thermal dependencies of $LIR_4$ and $LIR_5$-e) and corresponding $S_R$ - f).

As demonstrated above, the use of different spectral regions enables the development of four distinct ratiometric luminescent thermometers based on the emission spectra of $LaGaO_3$:0.001%$Mn^{4+}$, 0.25%$Tb^{3+}$. A comparison of the temperature-dependent $S_R$ for each approach reveals that the optimal operating temperature ranges for the individual thermometers are distinctly shifted with respect to one another. In this work, the optimal operating range of a thermometer is defined arbitrarily as the temperature interval within which the relative sensitivity exceeds 1% $K^{-1}$. Among the four, the thermometer based on $LIR_4$ demonstrates the broadest operational range, extending from approximately 200 K to 543 K. For lower temperatures, reliable readings can be achieved using $LIR_3$, which operates effectively between 170 K and 260 K. At even lower temperatures, $LIR_1$ offers accurate thermometric performance in the range of 100 K to 180 K. Consequently, by employing these complementary thermometric strategies, $LaGaO_3$:0.001%$Mn^{4+}$, 0.25%$Tb^{3+}$ enables luminescence thermometry across a remarkably wide temperature window of 100 - 543 K. From a practical standpoint, an essential figure of merit for thermometric performance is the uncertainty in temperature determination ($\delta T$). This value was calculated for each thermometer using the standard approach described by the following equation:

$$\delta T = \frac{1}{S_R} \frac{\delta LIR}{LIR} \quad (9)$$

where $\delta LIR/LIR$ represents the uncertainty in the $LIR$ determination, primarily influenced by the signal-to-noise ratio:

$$\frac{\delta LIR}{LIR} = \sqrt{\left(\frac{\delta I_1}{I_1}\right)^2 + \left(\frac{\delta I_2}{I_2}\right)^2} \quad (10)$$

where the $\delta I$ is the uncertainty of the emission intensity determination and was calculated as a integral emission intensity of the baseline (spectral range where no emission of was noticed) and $I_1$, $I_2$ represents the first and the second emission intensity used for LIR calculations, respectively. The lowest $\delta T$ was observed for $LIR_4$, reaching as low as 0.1 K in the temperature interval of 350 K - 390 K. Higher uncertainties were obtained for the other $LIR$-based thermometers, primarily due to their lower relative sensitivities and reduced emission intensities within their respective operating ranges. A distinctive feature of luminescent thermometers that rely on a first-order structural phase transition is the presence of thermal hysteresis in the $LIR$ versus temperature relationship. This hysteresis, arising from the difference in phase transition temperatures during heating and cooling cycles - as confirmed by DSC measurements - can affect the accuracy and reliability of temperature readings. To examine this effect, the thermal dependence of $LIR_4$ was measured during both heating and cooling cycles in the temperature range corresponding to the phase transition in $LaGaO_3$:0.001%$Mn^{4+}$, 0.25%$Tb^{3+}$. The results confirmed the presence of a minor but noticeable hysteresis loop. To mitigate this issue and improve measurement consistency, a modified calibration curve was proposed, denoted as $LIR_{5avr}$, which represents the arithmetic average of $LIR_4$ values obtained during heating and cooling:

$$LIR_{5avr}(T) = \frac{LIR_{5H}(T) + LIR_{5C}(T)}{2} \quad (11)$$

This averaged approach significantly reduces temperature determination uncertainty. The thermal dependence of $LIR_5$ is presented in Figure 6e. The corresponding uncertainty in temperature determination calculated from this curve was notably improved, reaching approximately 0.39 K - much lower than that derived from the original $LIR_5$ data.

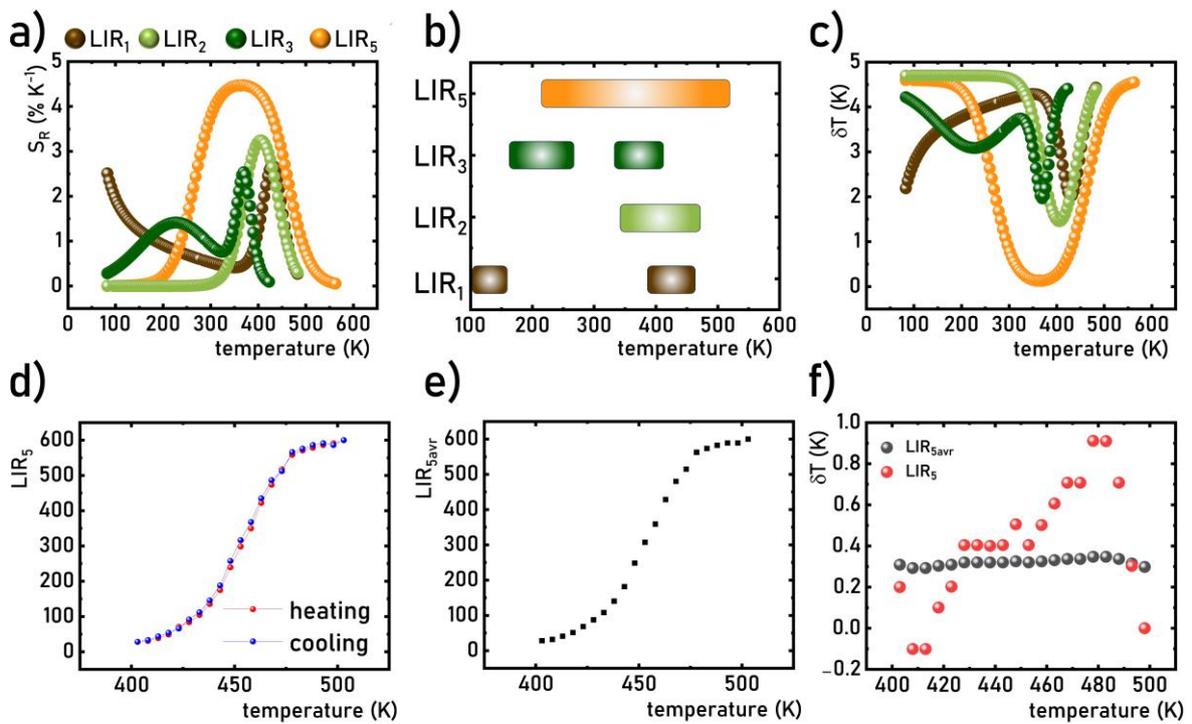

**Figure 6**. The comparison of thermal dependencies of $S_R$ for $LaGaO_3$:0.01%$Mn^{4+}$, $Tb^{3+}$ for different ratiometric approaches – a); the operating thermal ranges of these ratiometric approaches for which $S_R$>1 % $K^{-1}$ -b);

temperature determination uncertainties $\delta T$ for different ratiometric approaches for LaGaO$_3$:Mn$^{4+}$,Tb$^{3+}$ -c); thermal hysteresis of $LIR_5$ -d); thermal dependence of the $LIR_{5avr}$ -e); hysteresis induced temperature determination uncertainties resulted from $LIR_5$ and $LIR_{5avr}$ -f).

**Conclusions**

In this study, a novel approach to ratiometric luminescence thermometry is presented, based on the synergistic interplay of two thermally induced phenomena: (1) efficient thermal quenching of Mn$^{4+}$ luminescence associated with the $^2E \rightarrow {}^4A_2$ electronic transition, and (2) a first-order structural phase transition in LaGaO$_3$. As demonstrated, the structural transition from an orthorhombic to a rhombohedral phase near 420 K alters the local environment of the dopant ions, thereby modifying their spectroscopic properties. For Tb$^{3+}$ ions, this structural transformation leads to shifts in the Stark components of the emission bands, resulting in a thermally enhanced emission intensity in the high-temperature (HT) phase of LaGaO$_3$:Tb$^{3+}$. Exploiting the opposite thermal monotonicity of the Mn$^{4+}$ ions and Tb$^{3+}$ emission intensities in the HT phase of LaGaO$_3$ allows for a significant improvement in thermometric performance compared to the conventional approach based on the total integral intensity ratio of Mn$^{4+}$ and Tb$^{3+}$ emissions. Specifically, the phase transition-assisted strategy enables a maximum relative sensitivity of 4.7% K$^{-1}$ and extends the high-sensitivity operating range of luminescence thermometer to 220 K - 500 K. In contrast, the conventional $LIR$-based method yields a lower $S_R$ of 3.2% K$^{-1}$, with high sensitivity in the 350 K - 480 K range. Furthermore, it was observed that at very low Mn$^{4+}$ concentrations (0.001%), the HT phase of LaGaO$_3$:Mn$^{4+}$ exhibits notable spectral changes in respect to LT phase attributed to variations in crystal field strength and covalency of Mn$^{4+}$–O$^{2-}$ bonds. This enabled the development of a ratiometric thermometer based on the emission intensity ratio between HT and LT phases of LaGaO$_3$:Mn$^{4+}$, achieving a sensitivity of 2.7% K$^{-1}$ at 415 K over a narrow temperature range. However, this spectral behavior was not observed at higher Mn$^{4+}$ concentrations, likely due to the diminished emission intensity in the HT phase caused by enhanced interionic interactions. In addition, phase-transition-induced modifications in the $Tb^{3+}$ emission profile enabled the construction of a ratiometric thermometer based on the *LIR* between HT and LT phases of $LaGaO_3:Tb^{3+}$, achieving a maximum sensitivity of 2.8% $K^{-1}$ at 410 K. Importantly, since these effects occur simultaneously in $LaGaO_3$ co-doped with $Mn^{4+}$ and $Tb^{3+}$, the material functions as a multimodal ratiometric luminescence thermometer. Owing to the complementary thermal operating ranges of the individual thermometric modes, $LaGaO_3:Mn^{4+},Tb^{3+}$ enables temperature sensing across an exceptionally wide range, from 83 K to 503 K, while maintaining a sensitivity above 1% $K^{-1}$ throughout.

Overall, the results highlight the considerable potential of this phase-transition-assisted, multimodal thermometric strategy as a high-performance alternative to conventional ratiometric thermometers based solely on emission intensity ratios of transition metal and lanthanide ions.

**Acknowledgements**

This work was supported by the National Science Center (NCN) Poland under project no. DEC-UMO- 2020/37/B/ST5/00164.